\documentclass{ws-p9-75x6-50}

\newcommand{\bi}{\bibitem}

\newcommand{\rar}{\rightarrow}
\newcommand{\lrar}{\leftrightarrow}
\newcommand{\nue}{\nu_e}         
\newcommand{\num}{\nu_\mu}
\newcommand{\nut}{\nu_\tau}
\newcommand{\nus}{\nu_s}  
\newcommand{\dm}{\delta m^2}  

\begin{document}

\title{
Neutrinos and Big Bang Nucleosynthesis
}
\author{ A. D. Dolgov
}

\address{
INFN, sezione di Ferrara, Via Paradiso, 12 - 44100, Ferrara,
Italy;\\
ITEP, Bol. Cheremushkinskaya 25, 117218, Moscow, Russia
\\E-mail: dolgov@fe.infn.it}


\maketitle

\abstracts{
The role of neutrinos in big bang nucleosynthesis is discussed. The bounds
on the number of neutrino families, neutrino degeneracy, parameters of
neutrino oscillations are presented. A model of chemically inhomogeneous,
while energetically smooth, universe created by inhomogeneous cosmological
neutrino asymmetry is described. Nucleosynthesis limits on production of
right-handed neutrinos are considered.
}

\section{Introduction}

Big Bang Nucleosynthesis (BBN) is one of the strongest evidences in favor
of the Standard Hot Cosmological Model. The theory predicts abundances
of light elements $^2H$, $^3He$, $^4 He$, and $^7 Li$ which span
9 orders of magnitude in a good agreement with observations. 
Theoretical calculations are well defined and with the existing uncertainties
in the cross-sections of the relevant nuclear reactions the accuracy of
calculations is about 10\% or, depending upon the element, is even better,
especially for $^4 He$. However, comparison of theoretical results with 
observations is not straightforward because the data are subject to poorly
known evolutionary effects and systematic errors. Still, even with these
uncertainties, BBN permits to eliminate many modifications of the
standard model and to derive strong restrictions on properties of elementary
particles, in particular, on neutrino.

In what follows I will briefly discuss physics of BBN and essential
parameters that determine production of light elements (sec. 2). After
that the role of neutrinos in BBN is described. The bound on the number
of neutrino species is presented in section 3. In section 4 neutrino
degeneracy and its impact on BBN are discussed.
Neutrino oscillations and their possible influence on BBN are considered
in sec. 5. Inhomogeneous BBN related to possible spatial variations of
lepton asymmetry and producing large fluctuations of primordial abundances
in different regions of the universe is discussed in sec. 6.
In sec. 7 limits on some other neutrino properties (mass,
magnetic moments, right-handed currents, etc) are presented. 

The frameworks of the talk do not permit to give the
complete list of references
to the discussed problems, so I indicated only original and most recent
papers. A history of the problems and much more detailed list of references
can be found in the review~\cite{dolgov02}.

\section{Physics of BBN}
Building blocks for production of light nuclei were prepared in the
universe when the temperature was about 1 MeV. The reactions
\bea
n + e^+ &\lrar& p + \bar \nu_e, \nonumber \\
n +\nu_e &\lrar& p + e^-
\label{np-react}
\eea
maintained thermally equilibrium ratio of neutrons to protons
\be 
n/p = \exp (-\Delta m /T -\xi_e)
\label{npeq}
\ee
till the temperature $T$ dropped down below
$T_f=0.6-0.7$ MeV. Here $\Delta m = 1.3 $ MeV is the neutron-proton mass 
difference and $\xi_e =\mu_e /T$ is the dimensionless chemical potential of
electronic neutrinos. At temperature below $T_f$ the reactions (\ref{np-react})
became practically frozen, their rate $\Gamma \sim G_F^2 T^5$
became much smaller than the universe expansion rate, $H=\dot a/a\sim T^2$,
where $G_F$ is the Fermi coupling constant, $H$ is the Hubble parameter and 
$a(t)$ is the cosmological scale factor. Correspondingly the ratio $n/p$
would remain constant if neutrons were stable. Due to neutron decay with
the life-time $\tau_n = 888 \pm 2$ sec the $n/p$-ratio drops as
$\exp (-t/\tau_n)$.

Formation of light elements through the chain of reactions 
$p+n \rar\, {^2H} +\gamma$, $p+{^2H} \rar\, {^3 He}$, $n+{^2H} \rar\, {^3H}$, 
$^3He +n \rar {^4He}$, etc
started at $T=T_{NS} = 60-70$ keV. The concrete value of $T_{NS}$ depends
upon baryon-to-photon ratio $\eta = n_B /n_\gamma \approx 5\cdot 10^{-10}$
(found from BBN itself and now from the angular spectrum of CMBR as well)
and the nuclear binding energies. The temperature $T_{NS}$ is so much
smaller than the nuclear binding energy because of very large number  
of cosmic photons per one baryon.
At $T=T_{NS}$ all neutrons were quickly
captured and no free neutrons remained in the plasma. Since $^4He$ has the
largest binding energy practically all the neutrons, that survived to the
moment when $T=T_{NS}$, ended their lives in $^4He$ nuclei.

The universe cooling rate can be obtained if one compares 
the cosmological energy density expressed in terms of 
the Hubble parameter and the energy density of thermally equilibrium plasma:
\be
\rho_{tot} = {3H^2 m_{Pl}^2 \over 8\pi} =
{\pi^2\over 30} g_* T^4
\label{rhotot}
\ee
where $g_*$ counts the number of relativistic species in the primeval plasma;
it includes 2 from photons, $4\cdot (7/8)$ from $e^\pm$-pairs, and 
$2\cdot 3\cdot (7/8)$ from three light neutrino families:
\be
g_* = 10.75 + {7\over 4} \left( N_\nu -3 \right),
\label{gstar}
\ee
and the last term describes contribution of some non-standard energy. The 
latter could be additional neutrino species, possibly sterile, non-coupled to
$W$ and $Z$ bosons, or new abundant massive particles, or some unknown
form of dark energy (e.g. vacuum energy, or in other words, cosmological
constant). In the last two cases equation of state of these unknown forms
of matter would be different from relativistic equation of state $p=\rho/3$
valid for light neutrinos and thus the parameter $N_\nu$ could be a function
of time. Moreover, the effect of this parameter on production of different 
elements may differ from the effect induced by additional neutrinos.

As follows from eq.~(\ref{rhotot})
the universe cooling rate is determined by the expression:
\be
\left({t\over {\rm sec}}\right )\left({T\over {\rm MeV}}\right)^2 
= 0.74 \left( {10.75 \over g_*}\right)^2.
\label{tT2}
\ee
Since the cooling rate depends upon the effective number of particle species,
BBN is sensitive to any form of energy present in the cosmic plasma 
in the range of temperatures from a few MeV
down to 60 keV.

As one can see from the discussion above primordial abundances of light
elements are the functions of:
\begin{enumerate}
\item{}
weak interaction rate, which determines the moment when the reactions of 
neutron-proton transformation freezes; this rates is determined by the
neutron life-time;
\item{}
cosmological energy density parametrized as $(N_\nu -3)$;
\item{}
number density of baryons, $\eta_{10}= 10^{10}n_B/n_\gamma$;
\item{}
neutrino degeneracy, given by dimensionless chemical potentials,
$\xi_a =\mu_a/T$, where $a=e,\mu,\tau$; $\xi_a$ remains constant during 
adiabatic expansion if corresponding leptonic charge is conserved.

\end{enumerate}

\section{Number of neutrino families}
As we have already mentioned in the previous section, an addition of 
an extra neutrino family would change the freezing temperature of
$n/p$-transformation (\ref{np-react}). Comparing reaction rate and 
the rate of cosmological expansion one can find $T_f \sim g_*^{-1/6}$.
So with rising number of species the number density of neutrons and
correspondingly the mass fraction of produced $^4 He$ also rises.
Another effect leading in the same direction is that the time when
the nucleosynthesis temperature, $T_{NS}$ is reached also depends upon
$g_*$; according to eq.~(\ref{tT2}) $t_{NS} \sim g_*^{-1/2}$, so with
larger $g_*$ less neutrons would decay before the onset of nucleosynthesis
and more $^4 He$ would be produced. Both these effects give rise to
the increase of mass fraction of primordial helium-4 by approximately 
5\% if one extra neutrino family is added.

The dependence of helium production on the number of neutrino families
was first mentioned in ref.~\cite{hoyle64} and a little later in 
ref.~\cite{peebles66}. A detailed investigation was done in the
papers~\cite{shvartsman69,steigman77}. 

Comparing observational data with the theory one can deduce an upper limit
on the number of additional neutrino families, $\Delta N_\nu = N_\nu -3$.
According to the numerous 
literature on the subject, this limit oscillates between 2 and 0.1. The
recent analysis of ref.~\cite{olive02} gives $N_\nu = 1.8 - 3.9$.

It is usually assumed that neutrinos have the equilibrium spectrum:
\be
f_\nu = \left[\exp(E/T - \xi) +1\right]^{-1}
\label{fnu}
\ee
and the corresponding energy density of neutrinos plus antineutrinos is
\bea
\rho_\nu +\rho_{\bar\nu} &=&
{1\over 2\pi^2} \int_0^\infty dp p^3 \left[{1\over e^{E/T -\xi}+1} +
{1\over e^{E/T +\xi}+1}\right]  \nonumber \\
&=&{7\over 8}\,{\pi^2 T^4 \over 15}\left[1  +
{30\over 7}\,\left({\xi \over \pi}\right)^2 +
{15\over 7}\,\left({\xi \over \pi}\right)^4 \right]
\label{rhoxi}
\eea

The limit on the number of neutrino families presented above is obtained
under assumption that neutrinos are not degenerate, i.e. $\xi =0$. This limit
is also valid for non-equilibrium $\num$ and $\nut$ with the same energy 
density as the equilibrium ones. This is not so for electronic neutrinos
because the latter effect the $n/p$-ratio not only by their energy
density but also more directly by their spectrum, since they took part
in the reactions (\ref{np-react}). Surprisingly the spectrum of all neutrinos
noticeably deviates form the equilibrium one, though they may be exactly
massless. This deviation is induced by different temperatures of electrons
and neutrinos due to $e^+e^-$-annihilation after neutrino 
decoupling~\cite{dolgov92,dodelson92}. According to analytical estimate of 
ref.~\cite{dolgov92} the spectral distortion has the form:
\be
{\delta f_{\nue} \over f_{\nue} }\approx 3\cdot 10^{-4} \,\,{E\over T}
\left( {11 E \over 4T } - 3\right)
\label{dff}
\ee
where $\delta f = f - f^{(eq)}$.
The distortion of the spectra of $\num$ and $\nut$ is approximately twice
weaker. The shift of helium-4 mass fraction due to this neutrino heating
by the residual $e^+e^-$-annihilation is only a few$\times 10^{-4}$, though
the total neutrino energy density becomes larger than the standard one by 
about 3\%~\cite{dolgov97}.  There is
another effect of the similar magnitude and sign~\cite{heckler94}, namely
finite-temperature electromagnetic corrections to the energy density
of $\gamma e^+ e^-$-plasma. It adds 0.01 effective number of extra
neutrino species.

\section{Lepton Asymmetry}

If the number density of particles is different from the number density
of antiparticles, i.e. charge asymmetry is non-vanishing, it is described by 
a non-zero chemical potential, $\xi$ (\ref{fnu}). Of course this
description is valid only in the case of kinetic equilibrium when the
distribution in energy has the canonical form dictated by kinetic equation
with strong elastic scattering term. Fast annihilation processes imply
also opposite values of chemical potentials for particles and 
antiparticles: $\bar \xi = - \xi$. 

According to eq.~(\ref{fnu}) energy density of degenerate neutrinos in
thermal equilibrium is larger than the energy density of non-degenerate ones.
The role of degenerate $\num$ and $\nut$ in BBN is simply to increase the
total energy density which corresponds to 
\be
\Delta N_\nu  (\xi)= {15\over 7} \sum_{a=\mu,\tau}
\left[ \left(\xi_a \over \pi \right)^4
+  2\left(\xi_a \over \pi \right)^2\right]
\label{deltan-xi}
\ee
The bound on $\Delta N_\nu$ presented above can be translated to 
the bound on the magnitude of the chemical potentials. In particular, if
$\Delta N_\nu <1$ then $|\xi_{\mu,\tau}| < 1.5$. 

The bound on the value of $|\xi_e|$ is much stronger because degeneracy
of electronic neutrinos exponentially shifts $n/p$-ratio, see 
eq.~(\ref{npeq}). The BBN bounds on chemical potentials would be somewhat
weaker if combined variation of all chemical potentials is allowed. In this
case a large value of $|\xi_{\mu,\tau}|$ may be compensated by a relatively
small and positive $\xi_e$~\cite{bianconi91}. Recent analysis of the 
work~\cite{hansen01} based on additional information extracted from the
measurements of angular fluctuations of CMBR permits to obtain the limits:
\be
-0.01 <\xi_{\nue} < 0.2,\,\,\,\, |\xi_{\num,\nut}| < 2.6
\label{limitsxi}
\ee
under assumptions that the primordial fraction of deuterium is
$D/H = (3.0 \pm 0.4)\cdot 10^{-5}$. More details and references can be
found in the review~\cite{dolgov02}.

These results can be further strengthen if there are neutrino oscillations
that mix $\nue$, $\num$, and $\nut$. In this case muonic or tauonic 
asymmetries would be transformed through oscillations into electronic
asymmetry. This problem was analyzed recently in ref.~\cite{dolgov02-a}
where it was shown that for large mixing angle solution for the solar 
neutrino anomaly and for $(\num-\nut)$-mixing deduced from the atmospheric
neutrino anomaly the bounds on chemical potentials of all flavors are 
approximately
\be
|\xi_a| < 0.1
\label{xia}
\ee

\section{Neutrino Oscillations and BBN}
Effects of neutrino oscillations on BBN are very much different if
only active neutrinos are mixed or mixing is allowed between active and
hypothetical sterile neutrinos. In the first case oscillations do not
create any deviation from the standard BBN results if neutrinos are in
thermal equilibrium with vanishing chemical potentials. Indeed, in this case
oscillations would not lead to any modification of the standard distribution 
functions of neutrinos and abundances of light elements would remain the same
as they were without oscillations. A noticeable effect would arise if 
neutrinos are strongly degenerate and different lepton asymmetries would
be redistributed by oscillations as discussed in the previous section.

Physics is much more interesting if there are oscillations between active
and sterile neutrinos. One evident effect is that oscillations would create
additional neutrino species leading to $N_\nu >3$. Second, 
oscillations may distort spectrum of electronic neutrinos. The sign of the
effect may be both positive or negative depending upon the form of spectral
distortion. Third, in the case of MSW-resonance oscillations between 
$\nue$ and $\nu_s$
could be more efficient than oscillation between antineutrinos or vice versa.
This would give rise to generation of lepton asymmetry in the sector
of active neutrinos and, in particular, of electronic charge asymmetry
which would have a strong impact on BBN.

Excitation of additional, sterile, degrees of freedom at BBN by neutrino
oscillations was considered in many papers starting from 1990 (a large list
of references can be found in~\cite{dolgov02}). A recent 
bound on the mixing parameters between $\nu_s$ and $\nue$ and $\nus$ and
$\num$ or $\nut$ respectively reads~\cite{dolgov00-nr}:
\bea
(\dm_{\nue\nus}/{\rm eV}^2) \sin^4 2\theta_{vac}^{\nue\nus} =
3.16\cdot 10^{-5} (g_*(T^{\nus}_{prod})/10.75)^3 (\Delta N_\nu)^2
\label{dmess2}\\
(\dm_{\num\nus}/{\rm eV}^2) \sin^4 2\theta_{vac}^{\num\nus} =
1.74\cdot 10^{-5} (g_*(T^{\nus}_{prod})/10.75)^3 (\Delta N_\nu)^2
\label{dmmuss2}
\eea
where the number of relativistic degrees of freedom $g_*$ is taken at the
temperature $T^{\nus}_{prod} $ at which sterile neutrinos are effectively 
produced:
\be
T^{\nus}_{prod} = (12,\,15)\, (3/y)^{1/3}\,
(\dm/{\rm eV}^2)^{1/6}\,\, {\rm MeV}
\label{tprodnus}
\ee
These bounds 
are valid only if $\Delta N_\nu <1$ and spectral distortion of $\nue$ is 
neglected.

The impact on BBN
of the distortion of the spectrum of $\nue$ by oscillations
was discussed e.g. in ref.~\cite{kirilova99}. According to this work
an analytical fit to the bound on the oscillation
parameters that follows from the consideration of primordial
$^4 He$ can be written as
\be
\delta m^2\,\left(\sin^2 2\theta \right)^4 \leq
1.5\cdot 10^{-9}\,\, {\rm eV}^2,\,\,\,
{\rm for}\,\,\, \delta m^2 < 10^{-7}\,{\rm eV}^2.
\label{dmsintheta}
\ee

The situation is much more complicated for a larger mass difference. In 
particular, for the mass difference $\dm \sim (1-100)$ eV$^2$ and a small
vacuum mixing angle, $\sin^2 2\theta <10^{-3}$, the resonance amplification
of lepton asymmetry in the sector of active neutrinos can take
place~\cite{foot96}. The effect was discussed in detail in the subsequent 
literature and now seems to be confirmed both numerically and analytically.
The impact of this phenomenon on BBN could be quite strong but no
simple analytical results have been presented. An example of calculation of
the effective neutrino number $\Delta N_\nu$ induced by the generation of
asymmetry in $\nue$ sector can be found in ref.~\cite{dibari00}.

\section{Spatial Variation of Primordial Abundances}

An interesting phenomenon would arise if lepton asymmetries were 
inhomogeneous and large during BBN.
In this case a large spatial variation of primordial
abundances of light elements should take place.
If the scale of variation is larger than
the mixing scale (galactic), then the observed abundances of light elements
would be different in different regions of the universe. A mechanism of
generation of large and inhomogeneous lepton asymmetry together with
a small and possibly homogeneous baryon asymmetry was considered long ago
in ref.~\cite{dolgov-lp} in the frameworks of Affleck and Dine 
lepto/baryogenesis scenario. 
Another model of creation of large and inhomogeneous
lepton asymmetry by resonance neutrino oscillations (similar to the discussed
in the previous section) was suggested recently in the
paper~\cite{dibari99}. 

To avoid large density perturbations generated by varying chemical potentials
it was assumed~\cite{dolgov99-p} that there exists symmetry
with respect to permutation of electron, muon and tauon asymmetries.
In the simplest version of such model 
electron asymmetry is small over 2/3 of the sky 
but muon or tauon asymmetries are large (of order unity). Abundances
of light elements are normal there, i.e. the mass fraction of primordial
$^4 He$ is
$Y_p \approx 0.25$ and
the deuterium-hydrogen ratio is $D/H = 3\cdot 10^{-5}$. Over 1/6 of the sky
where electron asymmetry is large and negative and other asymmetries are
small the primordial abundances are high, $Y_p \approx 0.5$ and
$D/H = 10\cdot 10^{-5}$, and over other
1/6 of the sky where electron asymmetry is
large and positive the abundances are low, $Y_p \approx 0.12$ and
$D/H = 1.5\cdot 10^{-5}$. 
In more complicated versions of the model
the probability distribution of the regions with normal, high, 
and low primordial abundances could be different. 

The characteristic scale of variation is not predicted by the model but from 
the upper limit on angular variation of the temperature of CMBR one may
conclude that the scale should be larger than a few hundreds 
Mpc~\cite{dolgov99-p}. Still some features in the angular spectrum
of CMBR may be observable, in particular, diffusion damping slope at high
$l$ could be different in different directions in the sky.

The model discussed above presents an example of cosmology with broken
Copernicus Principle: the universe is energetically smooth but strongly
chemically inhomogeneous. 

\section{Non-standard Neutrino Properties}

If neutrinos are massive (with Dirac mass) or possess right-current
interactions then in addition to the usual left-handed species right-handed
neutrino states might be present at BBN. Kinematical excitation of 
right-handed neutrino states by their Dirac mass was considered in many
papers (see the review~\cite{dolgov02}). The latest and the most accurate
work~\cite{fields95} presents the limits:
\bea
m_{\nu_\mu} \: \leq \:
\left\{
\begin{array}{lll}
&130~{\rm keV}, &
\;  \; T_{QCD}=100~{\rm MeV} \\
&120~{\rm keV}, &
\;  \; T_{QCD}= 200~{\rm MeV} \nonumber
\end{array}
\right. \, \nonumber \\
m_{\nu_\tau} \: \leq \:
\left\{
\begin{array}{lll}
&150~{\rm keV}, &
\;  \; T_{QCD}=100~{\rm MeV} \\
&140~{\rm keV}, &
\;  \; T_{QCD}= 200~{\rm MeV.}
\end{array}
\right.
\label{mnur03}
\eea
if $\Delta N_\nu < 0.3$,
while for $\Delta N_\nu < 1.0$, they are
\bea
m_{\nu_\mu} \: \leq \:
\left\{
\begin{array}{lll}
&310~{\rm keV}, &
\;  \; T_{QCD}=100~{\rm MeV} \\
&290~{\rm keV}, &
\;  \; T_{QCD}= 200~{\rm MeV} \nonumber
\end{array}
\right. \, \nonumber \\
m_{\nu_\tau} \: \leq \:
\left\{
\begin{array}{lll}
&370~{\rm keV}, &
\;  \; T_{QCD}=100~{\rm MeV} \\
&340~{\rm keV}, &
\;  \; T_{QCD}= 200~{\rm MeV.}
\end{array}
\right.
\label{mnur10}
\eea
These limits are much stronger than laboratory limits for $\nut$ mass
and comparable to the limit on ${\num}$ mass. They are applicable if the
neutrino life-time is longer than the characteristic time of nucleosynthesis
(of course, because of Gerstein-Zeldovich limit~\cite{gz}
the life-time should be shorter than the universe age). 
On the other hand, an interpretation
of neutrino anomalies in terms of neutrino oscillations demands very small
mass difference and together with the laboratory bound, 
$m_{\nue} < 3$ eV (see ref.~\cite{mnue}),  
they lead to the masses of $\num$ and $\nut$ 
much below the bounds (\ref{mnur03},\ref{mnur10}).

If neutrinos, in addition to left-handed currents, are also coupled to 
right-handed ones, then the right-handed degrees of freedom could be present
at BBN even if neutrinos are strictly massless. To avoid too many 
right-handed neutrinos the interactions with right-handed currents should
be sufficiently weak and decouple at high temperatures such that the
entropy dilution due to massive particle annihilation would diminish the
number density of right-handed neutrinos down to a safe for BBN value. 
In other words, the mass of right-handed intermediate bosons should be
sufficiently high. There is some disagreement in the literature about the 
the exact value of the lower bound on their mass, but roughly speaking it
should be larger than a few TeV (see review~\cite{dolgov02}).

Another source of production of right-handed neutrinos could be their
magnetic moment. There are two possible types of processes in the 
early universe where neutrino spin flip due to magnetic moment might take 
place: first, direct particle reactions, either quasi-elastic scattering,
$e^\pm + \nu_L \rar e^\pm + \nu_R$, and annihilation,
$e^-+e^+ \rar \nu_{L,R} +\bar \nu_{R,L}$, or the
plasmon decay, $\gamma_{pl} \rar \bar \nu_{L,R} + \nu_{R,L}$. The second
process is the classical spin rotation of neutrinos in large scale primordial 
magnetic fields that might have existed in the early universe. 
The former mechanism was
first considered in ref.~\cite{morgan81}, while the second one in
refs.~\cite{shapiro81,lynn81}.

Assuming that the extra number of neutrino species allowed by BBN is smaller
than 1, the authors of the most recent paper~\cite{ayala99} 
found the limit:
\be
\mu_\nu< 2.9 \cdot 10^{-10}\mu_B
\label{munufin}
\ee
for the case of reaction produced neutrinos.

Potentially stronger limit can be obtained from consideration of neutrino
spin-flip in cosmological magnetic fields at nucleosynthesis epoch.
Unfortunately the result depends upon poorly known field strength in the
early universe. With a reasonable assumption about the latter the upper
bound on magnetic moment of neutrinos could be much stronger than 
(\ref{munufin}), 
see discussion in the review~\cite{dolgov02}.


\end{document}